\documentclass[reprint, amsmath,amssymb,prm]{revtex4-2}
\usepackage{graphicx}
\usepackage{dcolumn}
\usepackage{bm}
\usepackage[dvipsnames]{xcolor}
\usepackage{natbib}

\begin{document}

\preprint{AIP/123-QED}

\title{Laser induced transition in Cerium Oxide thin film\\}

\author{Mousri Paul}
\affiliation{%
Saha Institute of Nuclear Physics, 1/AF Bidhannagar, Kolkata 700064, India\\
}%
\affiliation{
Homi Bhabha National Institute, BARC Training School Complex, Anushaktinagar, Mumbai 400094, India
}%
\email{mousri.paul@saha.ac.in}

\author{Sabyasachi Karmakar}%
\affiliation{%
Saha Institute of Nuclear Physics, 1/AF Bidhannagar, Kolkata 700064, India\\
}%
\affiliation{
Homi Bhabha National Institute, BARC Training School Complex, Anushaktinagar, Mumbai 400094, India
}%

\author{Jay Sharma}%
\affiliation{%
Saha Institute of Nuclear Physics, 1/AF Bidhannagar, Kolkata 700064, India\\
}%
\affiliation{
Homi Bhabha National Institute, BARC Training School Complex, Anushaktinagar, Mumbai 400094, India
}%


\author{Satyaban Bhunia}
\affiliation{%
Saha Institute of Nuclear Physics, 1/AF Bidhannagar, Kolkata 700064, India\\
}%

\author{Supratic Chakraborty}
\affiliation{%
Saha Institute of Nuclear Physics, 1/AF Bidhannagar, Kolkata 700064, India\\
}%

\date{\today}

\begin{abstract}

The complex interaction between the intrinsic and extrinsic state variables of strongly correlated insulator thin films is drawing interest as it shows memristive behavior that may be used in neuromorphic computing. In this study, we have observed that laser irradiation in cerium oxide thin film helps to make transition from metastable state. We study the transport properties of cerium oxide thin film which shows the metal-insulator transition. As for ceria the reduction of oxidation states depends upon oxygen vacancy defects which can be controlled by external ways due to which the transport property can vary. In this study, a decrease in the value of transition temperature under dark conditions was observed as laser-induced and we have explained this result by the generation of energy states observed from temperature-dependent photoluminescence (PL) study, oxygen vacancy defects, and polaron hopping within ceria.

\end{abstract}

\maketitle


\section{\label{sec:level1}Introduction}

Cerium oxide is one of the most interesting rare earth materials and has many applications such as photocatalysis, and electrocatalysis \cite{das2013cerium, li2019review, lord2021redox}, in ferroelectric devices \cite{yu2022ceo, yu2022ceo2}and as memristors for application in RRAM devices \cite{saha2023experimental, de2020formation}.  Having the ability to reduce its Ce$^{4+}$ states to Ce$^{3+}$ depending upon the oxygen vacancy defects within cerium oxide, it shows some interesting properties at different experimental conditions. The polaron formation due to the oxygen vacancy defects and diffusion via the hopping mechanism strongly affects the electrical conductivity of cerium oxide. Naik et. al. reported that due to the formation of Ce$^{3+}$ states, the electrons are localized at the 4f state, and upon increasing the temperature these electrons jump from Ce$^{3+}$ to Ce$^{4+}$ states via hopping mechanism \cite{naik1978small}. This temperature-dependent polaron hopping helps in oxygen vacancy migration resulting in the formation of conducting filament and thus resistive switching in ceria-based devices was observed \cite{paul2023temperature}. In previous studies, a strain-dependent oxygen vacancy formation was observed at high temperatures \cite{paul2024effect}. This formation of oxygen vacancies and the corresponding reduction of Ce$^{4+}$ states to Ce$^{3+}$ is one of the key factors that can control the electronic charge transport property of ceria. The stoichiometric form of ceria has a fluorite structure with a large band gap of 6 eV which reduces due to the presence of oxygen vacancies as point defects. Photoluminescence (PL) is a powerful method to get an idea about the formation of Ce$^{3+}$ within CeO$_{2}$. Previously photoinduced phenomena were observed in ceria where the change in PL spectra highly depends upon the reduction of Ce$^{4+}$ due to oxygen vacancy defects \cite{skorodumova2002quantum}. A cerium oxide single crystal exposed to ultraviolet laser light has a PL spectrum that varies with time and is dependent on the surrounding gaseous atmosphere as reported by  Mochizuki et al \cite{mochizuki2009photoluminescence}. 

In this paper, we report the peak splitting in temperature-dependent PL spectra and correlate this with the corresponding current-voltage characteristics of ceria. A hysteresis loop in the R-T curve after a complete heating and cooling cycle was observed in dark conditions which vanishes after being excited by laser light during transport measurement. These experimental results confirm the laser-induced metal-insulator transition in ceria thin film.

\section{\label{sec:level2}Experimental details}

Cerium oxide was deposited by standard reactive magnetron sputtering technique using 99.999\% pure Ce metal target on $p$-type silicon (100) substrate having the resistivity of 0.5-5 $\Omega$-cm. The deposition of 20 nm CeO$_{2}$ film was carried out in the presence of argon and oxygen plasma environment by maintaining the deposition pressure of 5mTorr at 45 Watt rf power. A low-temperature photoluminescence (PL) study was performed using a He-Cd laser line of wavelength $\lambda$ = 325 nm, photon energy = 3.81 eV. For electrical characteristics, the metal pad of diameter 100$\mu$m with 25$\mu$m distance from each other is fabricated using the optical lithography technique and then titanium/gold bilayer was deposited by electron-beam evaporation. The electrical characteristics were performed within a temperature range of 4-300 K (in both heating and cooling cycles) under dark and incident laser-excited conditions. For transport measurements, we have used the red diode laser light of $\lambda$ = 380 nm and the power of 5mWatt.

\section{Results and discussion}

\begin{figure}[h!]
\includegraphics[scale=0.35]{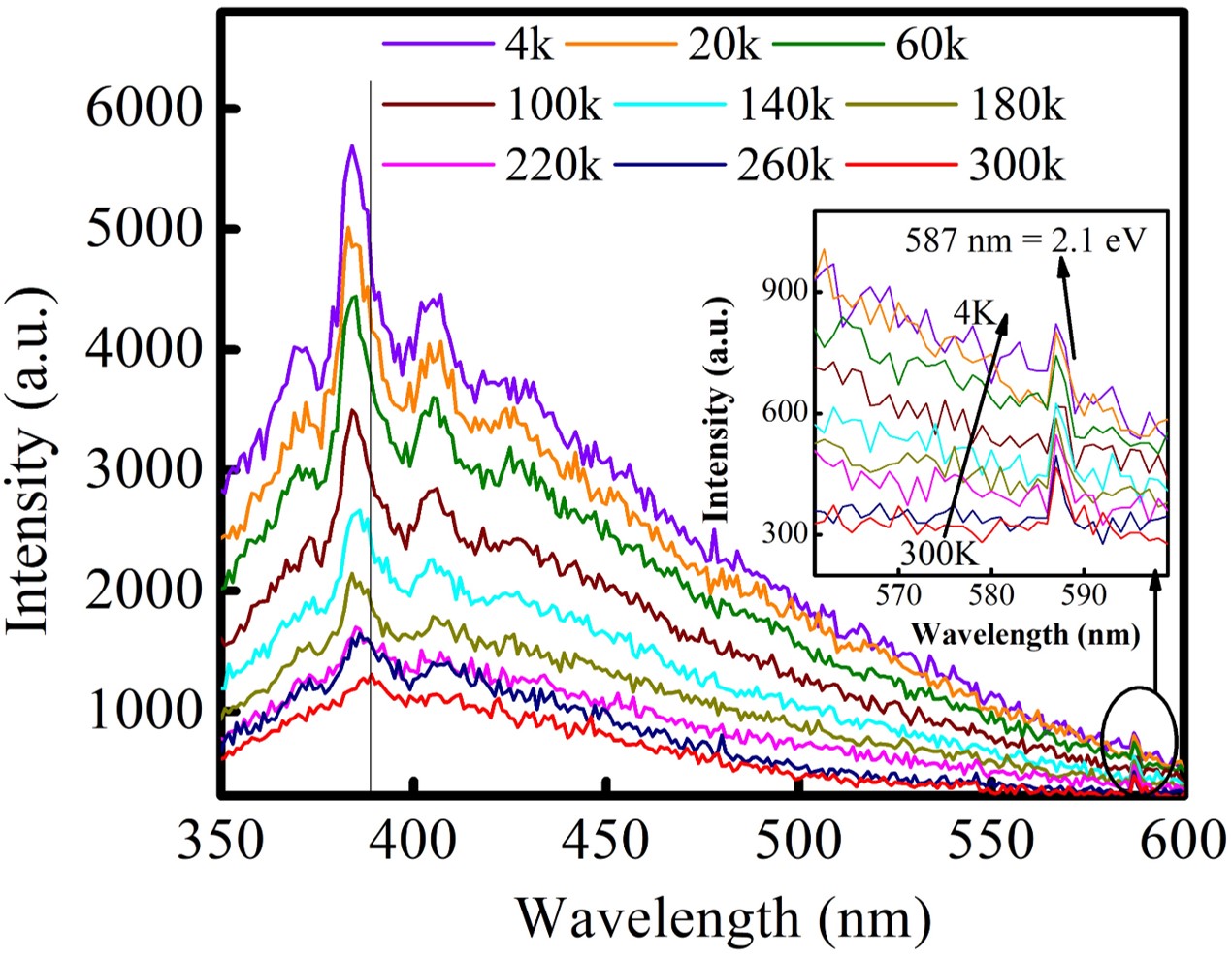}
\caption{\label{fig:pl} PL spectra of ceria thin film with temperature, inset corresponds to the zoomed view of the peak appeared at 587 nm (photon energy = 2.1 eV)}
\end{figure}

\begin{figure*}
\includegraphics[scale=0.55]{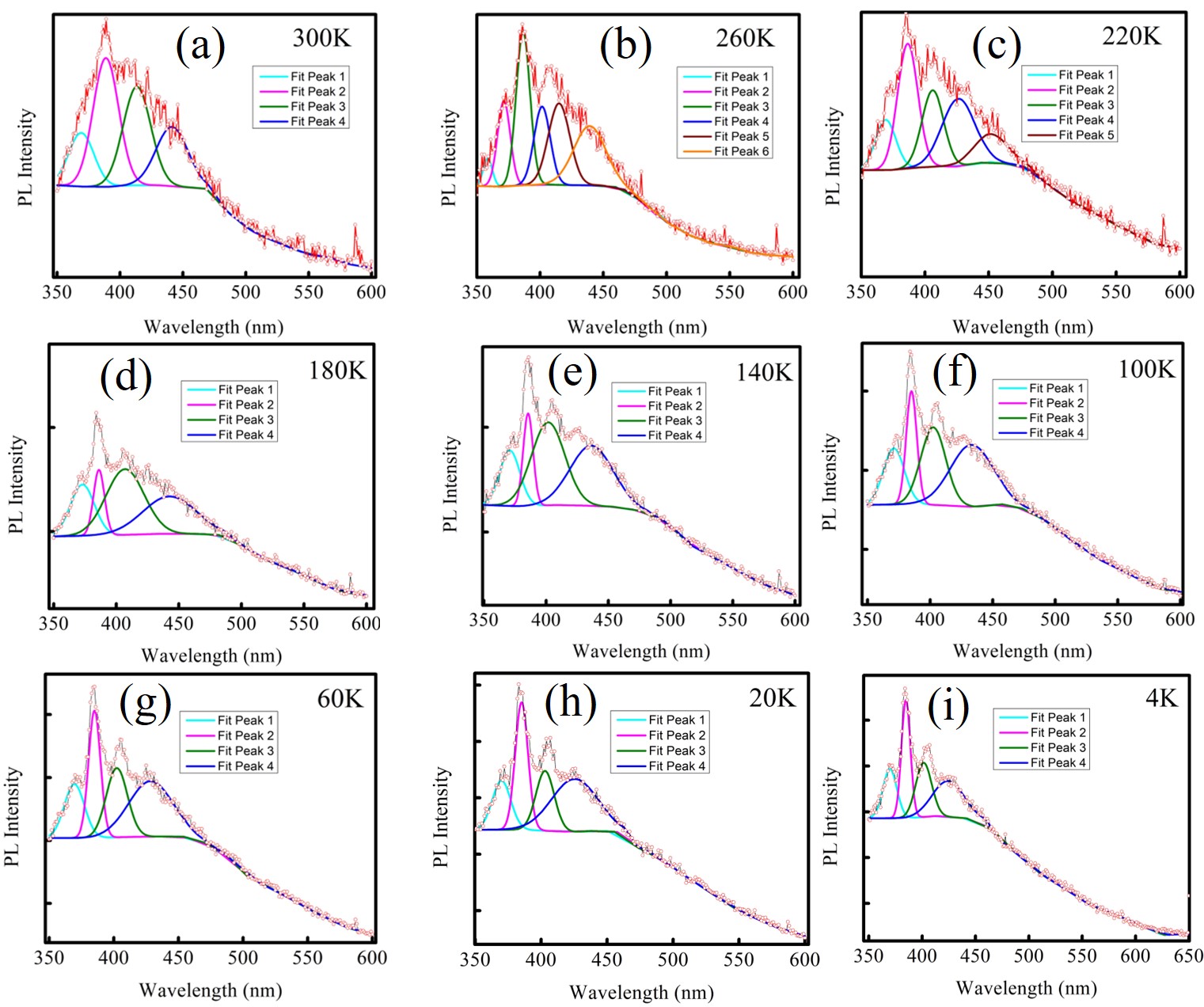}
\caption{\label{fig:pl_de} De-convoluted PL spectra of ceria thin film with temperature.}
\end{figure*}

Temperature dependence on the electron-lattice interaction was observed from the PL spectra of ceria and is presented in Fig \ref{fig:pl}. A major peak at the wavelength of 383 nm (peak I) with two distinct peaks (peak II and peak III) was observed at 4K and with increasing temperature the intensity of all three peaks decreased. As the temperature reaches 300K, all three peaks are merged and appear as a broad peak. One small peak at 587 nm was observed throughout this range of temperature. The deconvoluted peaks of PL spectra (for the 4-300K temperature range ) are presented in Fig. \ref{fig:pl_de}. It was also observed that the PL peak appears at the wavelength of 442 nm at 300K and vanishes as the temperature is decreased to 4 k.\\

The electrons are trapped in oxygen vacancy sites as the temperature reaches at 4K region and form the F centers (one trapped or two trapped electrons and without electrons)\cite{Morshed1997, choudhury2015annealing}. The UV emission peak denoted as peak I in Fig \ref{fig:pl} corresponds to the charge transfer from Ce4f to O2p state \cite{choudhury2012ce3+} at 4 k and this peak has a small shift with increasing temperature. At relatively higher temperatures, electrons become excited and localize to the 4f state of ceria due to electron-phonon interaction and Ce$^{3+}$ states can form\cite{tuller1977small}. The appearance of a broad peak at room temperature is attributed to the presence of oxygen vacancy defects which are acting as centers of radiative recombination for electrons to excite to 4f state of ceria \cite{deus2015photoluminescence, ganguly2013structural, mochizuki2003uv} are mostly present below the Ce4f level and highly dependent upon temperature \cite{morshed1997violet}. This presence of oxygen vacancies can reduce the fluorite stoichiometric form of ceria to a non-stoichiometric phase \cite{morshed1997violet}. The transition of the electron from 4f to 5d state of ceria was also observed as the peak arises at 3.4 eV at lower temperature regions \cite{nagai2023mechanisms}. 

 \begin{figure}[h!]
\includegraphics[scale=0.35]{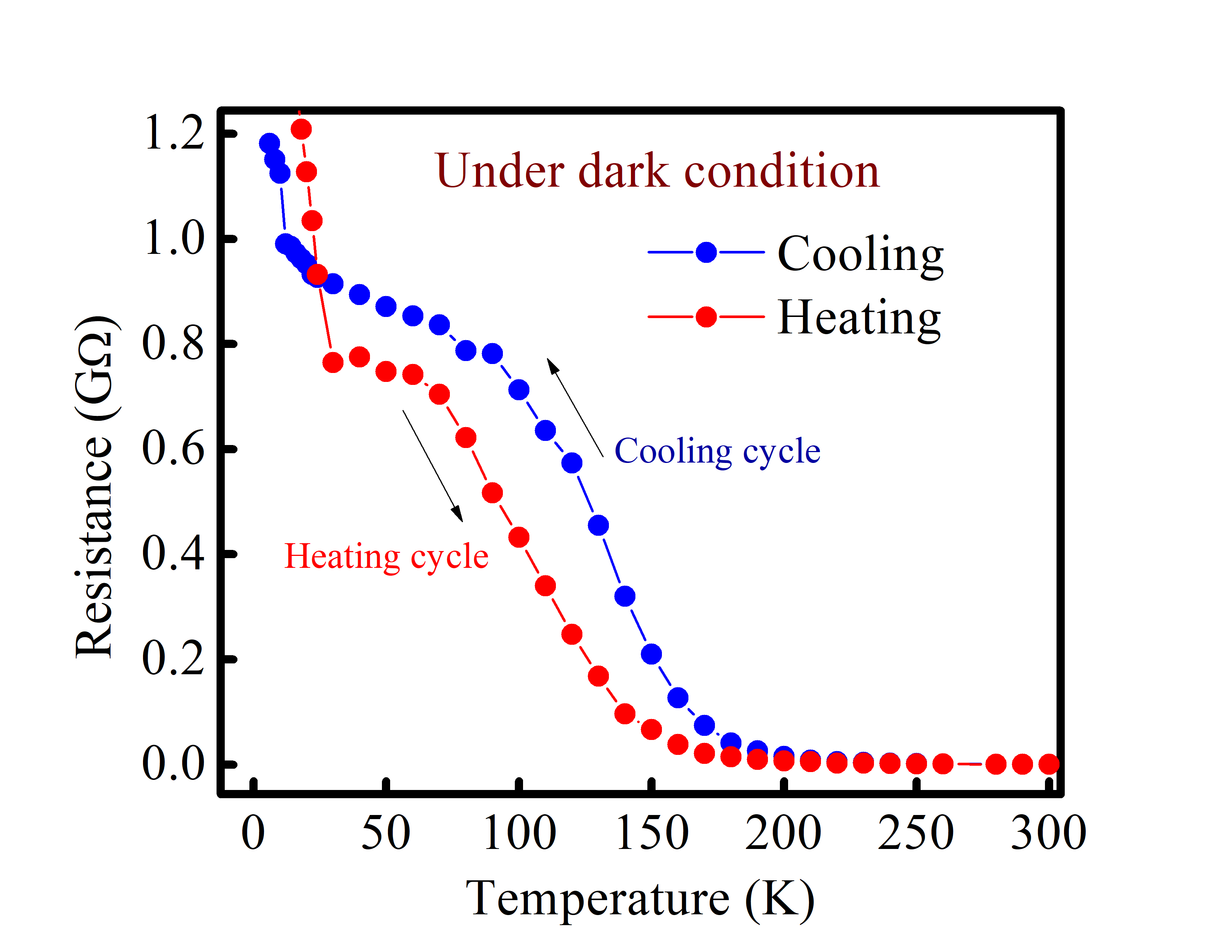}
\caption{\label{fig:iv} resistance vs temperature plot of cerium oxide thin film in dark condition}
\end{figure} 

 \begin{figure}
\includegraphics[scale=0.5]{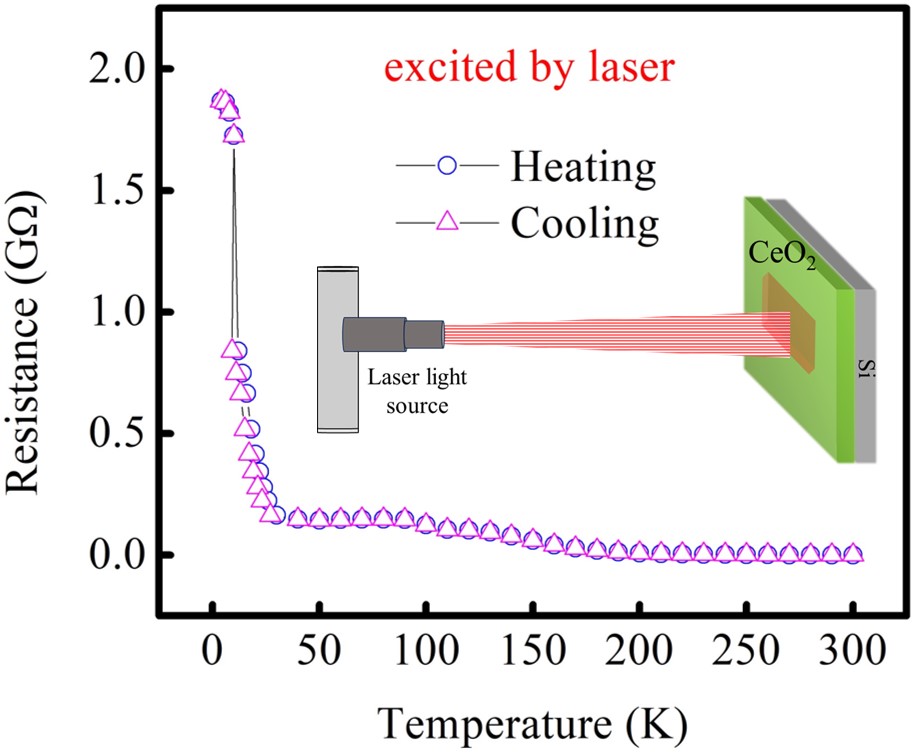}
\caption{\label{fig:laser1} resistance vs temperature plot of cerium oxide thin film by laser excitation}
\end{figure}

The current-voltage characteristics of ceria thin film were measured within the temperature range of 4k-300k under dark conditions and then measured under incident laser light. Fig. \ref{fig:iv} and Fig. \ref{fig:laser1} correspond to the resistivity vs temperature plot at a fixed voltage in both dark and laser excitation conditions of cerium oxide thin film. 
Very interestingly we have found that in the case of cerium oxide under dark conditions, a hysteresis occurs in the R(T) curve after a complete cooling and heating cycle, and it disappears upon applying the laser light. Under dark conditions, the R(T) curve gives a clear indication of the metal-insulator transition with a transition temperature of 150 K. Whereas, after exciting the sample by laser, a clear shift in the transition temperature towards the lower temperature region (nearly 30K) was observed.

In very recent work for vanadium oxide, the laser-induced quenching in Mott insulator-metal transition was observed where with increasing the laser intensity the hysteresis in R(T) curve was fully diminished, confirming the enhancement of relaxation of the system from a metastable state by laser irradiation\cite{luibrand2024laser}. In our case, we have also seen the same result in the R(T) curve measured in the laser-excited conditions and in this paper, we tried to explain the transport property with the corresponding PL results.
 
The reduction of CeO$_{2}$ due to oxygen vacancies can form a point defect, complex defects Ce$^{3+}$-O$_{v}$-Ce$^{4+}$, or O$_{v}$+2e$^{-}$\cite{lu2006morphology, norenberg1997defect}. The well-known band gap of CeO$_{2}$ is 6 eV between Ce 5d and O 2p states. The change in the oxidation state of ceria due to oxygen vacancies can be described in terms of band structure where one electron localizes at the Ce4f state and Ce$^{3+}$ can form. As the 4f state of ceria appears, the band gap decreases to 3 eV (band gap between O 2p and Ce 4f state)\cite{nagai2023mechanisms} and this state becomes narrower as compared to O2p states \cite{hay2006theoretical}. Therefore, an electron can trap within the Ce4f band having a larger effective mass with lower mobility, resulting in decreasing the electrical conductivity. 

When the laser light of wavelength 325 nm incident on the film, it introduces a p$\rightarrow$f interband transition resulting in the formation of photogenerated electrons (e) and holes (h). This photogenerated e and h in the phonon field was determined by three different energies proposed by Toyozawa theoretically \cite{toyozawa1983symmetry} such as the charge carrier and lattice relaxation energy, the Coulomb interaction energy U between e and h, and the kinetic energy of electronic transition. 


The formation energy of oxygen vacancy was theoretically calculated by Hersland et. al. within the range of 3.1-3.3 eV for Ceria (100) \cite{herschend2005electronic}. By considering the calculated value of formation energy as correct, the 325 nm laser photon energy can directly induce the oxygen vacancy defects within ceria resulting in strong localization of electrons in the Ce4f state, enhancing the electron-lattice interactions and it is regarded as a self-trapped exciton in the corresponding PL spectra \cite{mochizuki2009photoluminescence}. This formation of polarons is highly dependent on temperature.  From the experimental value of PL spectra of cerium oxide we have seen that at lower temperatures, two states around 3.0 eV appeared at 2.8 and 3.2 eV, which we consider as near band edge emission associated with many phonon emissions.  
 
As temperature decreases the thermal energy decreases and the polarons require more energy to hop, resulting in a decrease in the electrical conductivity of ceria. Therefore, under dark conditions as the temperature goes below 140K, the resistance starts to increase. By introducing laser light, the polarons get that much energy to hop to the conduction band at a lower temperature region, and as the temperature goes below 30K, the self-trapped polarons have a higher relaxation time, resulting in a decrease in the electrical conductivity of the film. The self-trapped polarons also contribute to PL spectra at lower temperatures by introducing oxygen vacancy defects where the defects act as non-radiative centers. As a result, the excited electrons can trap, and the carrier mobility decreases. Whereas, with increasing the temperature the thermal energy was introduced and the trapped electrons are getting energy to overcome the barrier and contribute to the conductivity of ceria. 
The combined results of the PL and I-V study confirm the metal-insulator transition temperature in cerium oxide which is highly affected by laser illumination. 

\section{Conclusion}

We have explained the effect of laser induction on metal-insulator transition temperature in cerium oxide thin film and correlated the results with corresponding PL spectra. The oxygen vacancy defect formation due to the laser induction can strongly localize the electrons in the Ce 4f state. Under room temperature conditions different types of emission can occur resulting in broadening of PL spectra of ceria. By decreasing the temperature from 300K, the phonon energy becomes minimized, and as a result, only near-band edge emissions are dominant, resulting in the splitting of PL spectra. At lower temperature regions, the self-trapped electrons forming the small polarons present in the Ce 4f state, have larger effective mass with smaller mobility decreasing the electrical conductivity. The transition temperature of ceria under dark conditions was improved by introducing a small perturbation by laser excitation. This study confirms that for cerium oxide thin film the enhancement of relaxation from the metastable state was observed which can be useful for low temperature memristive device applications

\begin{acknowledgments}


M. Paul and S. Karmakar wish to acknowledge the financial assistance received from the University Grants Commission, Government of India.

\end{acknowledgments}

\nocite{*}

\bibliography{biblio1}
\bibliographystyle{unsrt}
\end{document}